\documentclass[11pt,a4paper]{article}
\usepackage{epsfig,amssymb,amsmath}
\newcommand{\news}{\setcounter{equation}{0}}
\newcommand{\be}{\begin{equation}}
\newcommand{\ee}{\end{equation}}
\newcommand{\bea}{\begin{eqnarray}}
\newcommand{\eea}{\end{eqnarray}}
\font\upright=cmu10 scaled\magstep1
\font\sans=cmss12
\newcommand{\ssf}{\sans}
\newcommand{\stroke}{\vrule height8pt width0.4pt depth-0.1pt}
\newcommand{\Z}{\hbox{\upright\rlap{\ssf Z}\kern 2.7pt {\ssf Z}}}

\newcommand{\C}{{\rlap{\rlap{C}\kern 3.8pt\stroke}\phantom{C}}}
\newcommand{\R}{\hbox{\upright\rlap{I}\kern 1.7pt R}}
\newcommand{\CP}{\C{\upright\rlap{I}\kern 1.5pt P}}
\newcommand{\identity}{{\upright\rlap{1}\kern 2.0pt 1}}
\newcommand{\half}{\frac{1}{2}}
\newcommand{\pr}{\partial}

\newcommand{\E}{{\cal E}}
\newcommand{\lam}{\lambda}

\begin{document} 
\title{\vskip -70pt \begin{flushright} {\normalsize
DAMTP-2008-85} \\ \end{flushright} 
\vskip 40pt 
{\bf{Scaling Identities for Solitons beyond Derrick's Theorem}}
\vskip 15pt} 
\author{{\bf Nicholas S. Manton}
\thanks{N.S.Manton@damtp.cam.ac.uk} \\ \\DAMTP, University of
Cambridge, \\ Wilberforce Road, Cambridge CB3 0WA, UK}
\maketitle 
\begin{abstract} 
New integral identities satisfied by topological solitons in a range
of classical field theories are presented. They are derived by
considering independent length rescalings in orthogonal directions,
or equivalently, from the conservation of the stress tensor. These
identities are refinements of Derrick's theorem. 
\end{abstract}

\section{Introduction}
We will be concerned here with static soliton solutions occurring in a
range of field theories in two, three or four dimensional
space. Solitons are smooth, localized, finite energy stationary points of
some energy function, either minima or saddle points \cite{ManSut}. The 
energy is the integral of a sum of terms
involving the field and its first partial derivatives. Our solitons 
are generally stabilised by some topological charge that prevents their
decay to the vacuum. Whereas the vacuum has zero
energy, a soliton has positive energy. Topological aspects of 
solitons will actually be unimportant here. However, it will be 
important that the field of a soliton rapidly approaches 
the vacuum field at spatial infinity, at least up to a gauge 
transformation, so that the energy density falls off rapidly too. 

It is well known that by rescaling lengths one can derive some
identities relating different contributions to the total energy of a 
soliton. The basic point, noted by Derrick \cite{De}, is that the energy at the
soliton solution is stationary with respect to a uniform spatial
rescaling (or dilation). Setting the variation with respect to an
infinitesimal dilation to zero gives non-trivial identities. In some
field theories, as Derrick showed, the scaling argument can be used to
prove that no finite energy solitons can exist. We shall not be
interested in this situation here.

In this letter we shall consider the effect of 
independent length rescalings in different Cartesian
directions. In various examples, we obtain simple and novel identities
connecting contributions to the total energy. Our results are
refinements of the Derrick scaling identities, in that simple
combinations of our results reproduce those of Derrick.

It is unlikely that many of our results are original, and in some sense
they are not new at all. From the conservation of the stress tensor
(the purely spatial analogue of the energy-momentum tensor), which is
a local identity that follows from the field equations, one can derive
an infinite number of integral identities, and ours are special 
cases, albeit interesting ones.

We shall not develop our results in a completely general way. Instead
we shall obtain scaling identities for Skyrmions, which are solitons 
in a purely scalar field theory in three dimensions, and then show how to
obtain scaling identities in theories with gauge fields coupled to 
scalar fields, whose solitons are vortices in two 
dimensions and monopoles in three dimensions. We shall 
also consider instantons, which are solitons in a pure gauge theory in four
dimensions.

In Section \ref{sec:Sky}, devoted to Skyrmions, we carry out
the scaling manipulations directly on the energy function. In Section 
\ref{sec:stress} we show how the same results are derivable from the
stress tensor of the field theory. Finally, in Section \ref{sec:gauge}
we introduce gauge fields coupled to scalar fields, and use the gauge
invariant stress tensor to obtain scaling identities for vortices and 
monopoles, and also for instantons and other solutions in pure gauge 
theories. Some solitons satisfy first order Bogomolny
equations. and for these solitons the stress tensor vanishes
identically, and our identities become trivial.

\section{Skyrmions}\news
\label{sec:Sky}

The Skyrme model in its original version is defined in three space
dimensions \cite{Sky}. We use Cartesian coordinates ${\bf x} 
= (x^1, x^2, x^3)$. The scalar field $U({\bf x})$ takes values in
$SU(2)$, so it is convenient to work not directly with the derivatives
of $U$ but with the currents $R_i = \pr_i U U^{-1}$, each component of
which takes values in the Lie algebra of $SU(2)$. The Skyrme energy
function, in simplified units, is
\be
\label{Skenergy}
E = \int \left\{ -\frac{1}{2}\,\hbox{Tr}\,(R_i R_i) 
- \frac{1}{16}\,\hbox{Tr}\,([R_i,R_j][R_i,R_j]) 
+ m^2\,\hbox{Tr}\,(1 - U) \right\} d^3 x \,.
\ee
Each term in $E$ is non-negative. What is important here is not so 
much the Lie algebra structure of the
commutators and traces, but which derivatives occur in each term.
Note that because of the commutators, no current component occurs
raised to the fourth power. $m^2\,\hbox{Tr}\,(1 - U)$ is the pion mass
term and is sometimes omitted.

A Skyrmion is a smooth, finite energy minimum or stationary point of
this energy, satisfying the boundary condition $U \to 1$ as $|{\bf x}|
\to \infty$. Skyrmions are distinguished by their topological charge,
their baryon number $B$, which is the degree of the map $U: \R^3 \to
SU(2)$. Many Skyrmions are known, at least numerically, for a large
range of values of $B$ \cite{BS}.

Suppose now that the field $U({\bf x})$ is a Skyrmion, and
consider an independent rescaling of the three coordinates:
\be
x^1 \to \lam_1 x^1 \,, \quad x^2 \to \lam_2 x^2 \,, \quad 
x^3 \to \lam_3 x^3 \,.
\ee
That is, we replace $U(x^1, x^2, x^3)$ by ${\widetilde U}(x^1, x^2, x^3)
= U(\lam_1 x^1, \lam_2 x^2, \lam_3 x^3)$. Clearly $\pr_i {\widetilde U}$
equals $\lam_i \pr_i U$. By changing variables, we can relate each term in
the energy ${\widetilde E}$ of the transformed field ${\widetilde U}({\bf x})$
to the corresponding term in $E$. The result is
\bea
{\widetilde E} &=& \int \left\{ 
-\frac{1}{2}\,\frac{\lam_1}{\lam_2\lam_3} \hbox{Tr}\,(R_1 R_1) 
-\frac{1}{2}\,\frac{\lam_2}{\lam_3\lam_1} \hbox{Tr}\,(R_2 R_2)
-\frac{1}{2}\,\frac{\lam_3}{\lam_1\lam_2} \hbox{Tr}\,(R_3 R_3) \right.
\nonumber \\
&& \quad 
- \frac{1}{8}\,\frac{\lam_1\lam_2}{\lam_3} \hbox{Tr}\,([R_1,R_2][R_1,R_2])
- \frac{1}{8}\,\frac{\lam_2\lam_3}{\lam_1} \hbox{Tr}\,([R_2,R_3][R_2,R_3])
\nonumber \\
&& \quad \left.
- \frac{1}{8}\,\frac{\lam_3\lam_1}{\lam_2} \hbox{Tr}\,([R_3,R_1][R_3,R_1]) 
+ \frac{1}{\lam_1\lam_2\lam_3} m^2\,\hbox{Tr}\,(1 - U) \right\} d^3 x \,,
\eea  
where the expressions on the right hand side are evaluated for the
field $U({\bf x})$. Because $U({\bf x})$ is a Skyrmion, the function 
${\widetilde E}$ is stationary with respect to $\lam_1$, $\lam_2$ and
$\lam_3$ at $\lam_1 = \lam_2 = \lam_3 = 1$. Rather than work out the general
consequences of this, it is simpler to restrict to special rescalings
(which together span all possibilities). Particularly worthwhile is to 
set $\lam_1 = \lam_2 = \lam$ and $\lam_3 = 1$. Then
\bea
{\widetilde E} &=& \int \left\{ 
-\frac{1}{2}\, \hbox{Tr}\,(R_1 R_1) 
-\frac{1}{2}\, \hbox{Tr}\,(R_2 R_2)
-\frac{1}{2}\,\frac{1}{\lam^2} \hbox{Tr}\,(R_3 R_3) \right.
\nonumber \\
&& \quad 
- \frac{1}{8}\, \lam^2 \hbox{Tr}\,([R_1,R_2][R_1,R_2])
- \frac{1}{8}\, \hbox{Tr}\,([R_2,R_3][R_2,R_3])
\nonumber \\
&& \quad
\left. - \frac{1}{8}\, \hbox{Tr}\,([R_3,R_1][R_3,R_1]) 
+ \frac{1}{\lam^2} m^2\,\hbox{Tr}\,(1 - U) \right\} d^3 x \,.
\eea
Notice that several terms are unaffected by this rescaling. The
derivative of ${\widetilde E}$ with respect to $\lam$ must vanish at $\lam =
1$, and this gives the identity
\be
\label{scaleid1}
\int \left\{ 
-\frac{1}{2}\, \hbox{Tr}\,(R_3 R_3)
+ \frac{1}{8}\, \hbox{Tr}\,([R_1,R_2][R_1,R_2])
+ m^2\,\hbox{Tr}\,(1 - U) \right\} d^3 x = 0 \,. 
\ee
Similarly, by permutation, one obtains
\be
\label{scaleid2}
\int \left\{ 
-\frac{1}{2}\, \hbox{Tr}\,(R_1 R_1)
+ \frac{1}{8}\, \hbox{Tr}\,([R_2,R_3][R_2,R_3])
+ m^2\,\hbox{Tr}\,(1 - U) \right\} d^3 x = 0 \,, 
\ee
\be
\label{scaleid3}
\int \left\{ 
-\frac{1}{2}\, \hbox{Tr}\,(R_2 R_2)
+ \frac{1}{8}\, \hbox{Tr}\,([R_3,R_1][R_3,R_1])
+ m^2\,\hbox{Tr}\,(1 - U) \right\} d^3 x = 0 \,. 
\ee
These identities appear to be novel, at least in the context of
Skyrmions. The sum of the three identities is
\be
\label{SkDerrick}
\int \left\{ -\frac{1}{2}\,\hbox{Tr}\,(R_i R_i) 
+ \frac{1}{16}\,\hbox{Tr}\,([R_i,R_j][R_i,R_j]) 
+ 3m^2\,\hbox{Tr}\,(1 - U) \right\} d^3 x = 0 \,,
\ee
and this is Derrick's theorem for Skyrmions, which is also obtained by
considering the uniform rescaling $\lam_1 = \lam_2 = \lam_3 = \lam$.

The difference of identities (\ref{scaleid2}) and (\ref{scaleid3}) is
\bea
&& \int \left\{ 
-\frac{1}{2}\, \hbox{Tr}\,(R_1 R_1)
+\frac{1}{2}\, \hbox{Tr}\,(R_2 R_2) \right. \nonumber \\
&& \quad \quad 
\left. - \frac{1}{8}\, \hbox{Tr}\,([R_1,R_3][R_1,R_3]) 
+ \frac{1}{8}\, \hbox{Tr}\,([R_2,R_3][R_2,R_3]) \right\} \, d^3x
= 0 \,, 
\eea
and rotating this by $45^\circ$ in the $(x^1, x^2)$ plane gives
\be
\label{scaleid4}
\int \left\{ 
-\frac{1}{2}\, \hbox{Tr}\,(R_1 R_2)
- \frac{1}{8}\, \hbox{Tr}\,([R_1,R_3][R_2,R_3]) \right\} \, d^3x
= 0 \,, 
\ee
which is a further novel identity. Two more identities are obtained by
permuting the indices:
\bea
\label{scaleid5}
&& \int \left\{ 
-\frac{1}{2}\, \hbox{Tr}\,(R_2 R_3)
- \frac{1}{8}\, \hbox{Tr}\,([R_2,R_1][R_3,R_1]) \right\}\, d^3x
= 0 \,, \\
\label{scaleid6}
&& \int \left\{ 
-\frac{1}{2}\, \hbox{Tr}\,(R_3 R_1)
- \frac{1}{8}\, \hbox{Tr}\,([R_3,R_2][R_1,R_2]) \right\} \, d^3x 
= 0 \,.
\eea

These identities could be useful in the following way. In the Skyrme
model, and in other field theories with solitons, one is often
interested in finding exact numerical solutions, by relaxation of the 
energy. One 
often starts with an analytical approximation to a solution, given by
simple formulae. For example, in the Skyrme model, the rational map ansatz
provides a good starting point \cite{HMS}. It is helpful to get
the size right by imposing the Derrick identity, and if it is not
initially satisfied, a simple uniform rescaling will ensure that it is.

With the new identities (\ref{scaleid1}) -- (\ref{scaleid3}) we can 
also see how to
improve a trial solution. For example, the rational map ansatz has no
squashing degrees of freedom, but now we can introduce them. Given a
field configuration, one should calculate the integrals
\bea
I_1 &=& \int \left\{ -\frac{1}{2}\,\hbox{Tr}\,(R_3 R_3)  
+ m^2\,\hbox{Tr}\,(1 - U) \right\} \, d^3 x \,,
\nonumber \\
I_2 &=&  \int \left\{
- \frac{1}{8}\,\hbox{Tr}\,([R_1,R_2][R_1,R_2]) \right\} \, d^3x \,. 
\eea 
Then a rescaling in the $(x^1, x^2)$ plane, with no
rescaling of $x^3$, by the factor 
\be
\lam = \left( {\frac{I_1}{I_2}} \right)^{\frac{1}{4}} 
\ee
will optimally reduce the energy. Following this one can rescale in,
say, the $(x^2, x^3)$ plane. This procedure can be iterated
until the optimal squashing is attained. The process converges, as the
energy cannot go below that of the true solution. Moreover, at each
step the integrals contributing to the energy change, but in a simple
way, so they do not need to be numerically calculated more than once.

This process could be particularly useful for finding good
approximations to Skyrmions with $B = 8, 10$ and $12$ (among
others). Here it is known that the usual rational map ansatz gives
trial solutions that are too spherical, whereas the true solutions are
rather prolate (for $B=8$), or oblate (for $B=10$ and $B=12$) 
\cite{BMS,BS1}. The cyclic symmetries of these solutions are a guide to which
rescalings are likely to be helpful.

It is possible that from a given starting field configuration, a single step
squashing/shearing transformation can be found, after which all six
identities (\ref{scaleid1}) -- (\ref{scaleid3}) and 
(\ref{scaleid4}) -- (\ref{scaleid6}) are satisfied, but 
we have not found an algorithm for this.

\section{Stress Tensor}\news
\label{sec:stress}

For a generic field theory in Euclidean space $\R^n$, with
translational invariance, one can derive using Noether's theorem a
conserved stress tensor. This is the static analogue of the
energy-momentum tensor of a dynamical field theory. If the energy
density is $\E$, and is a function just of some fields $\phi$ and
their spatial derivatives $\pr_i\phi$, then the stress tensor is
\be
\label{stress}
T_{ij} = \frac{\pr\E}{\pr(\pr_i\phi)} \pr_j\phi - \delta_{ij}\E \,.
\ee
The stress tensor satisfies the conservation law $\pr_i T_{ij} = 0$,
which can be verified by using the field equation
\be
\pr_i\left(\frac{\pr\E}{\pr(\pr_i\phi)}\right) - \frac{\pr\E}{\pr\phi}
= 0 \,.
\ee

Now let $V_j(\bf{x})$ be an arbitrary vector field (classical
functions, not a dynamical quantity) and define
\be
P_i = V_j T_{ij} \,.
\ee
Then $\pr_i P_i = (\pr_i V_j) T_{ij}$. Integrating this over $\R^n$,
and using the divergence theorem, gives
\be
\label{stressint1}
\int (\pr_i V_j) T_{ij} \, d^nx = 0 \,,
\ee
provided $P_i$ decays rapidly enough towards spatial infinity. We
shall consider field configurations $\phi$ which rapidly approach the vacuum
at infinity, so that $\E$ and $T_{ij}$ rapidly approach zero too. 
$V_i$ may grow towards infinity, but must not do so too fast.

We can rederive Derrick's theorem and our new scaling identities by
choosing $V_j$ to depend linearly on Cartesian coordinates. Let
\be
V_j = A_{jk} x^k
\ee
with $A_{jk}$ an arbitrary constant matrix. Then $\pr_i V_j = A_{ji}$,
so (\ref{stressint1}) implies that 
\be
\int T_{ij} \, d^nx = 0
\ee
for any labels $i,j$.

Let us see how this works for a Skyrmion. From the energy
expression (\ref{Skenergy}) we obtain, by applying (\ref{stress}), 
the stress tensor
\be
T_{ij} = -\hbox{Tr}\,(R_i R_j) 
- \frac{1}{4} \hbox{Tr}\,([R_i,R_k][R_j,R_k]) - \delta_{ij}\E \,.
\ee
The identity obtained by integrating $T_{11}$ gives a rather
complicated expression, but the combination
\be
\int (T_{11} + T_{22}) \, d^3x = 0
\ee
reproduces the identity (\ref{scaleid1}). The identities 
(\ref{scaleid2}) and (\ref{scaleid3})
are obtained similarly. The trace combination
\be
\int T_{ii}  \, d^3x = 0
\ee
is Derrick's theorem (\ref{SkDerrick}). The vanishing of the integrals of the 
off-diagonal elements of the stress tensor give the remaining
identities. For example, integrating $T_{12}$, one reproduces 
(\ref{scaleid4}).

For Skyrmions, there are no surface terms at infinity to worry 
about, because $1 - U$ decays as $|{\bf x}|^{-2}$ when $m=0$ 
\cite{Man}, so $R_i$ decays as $|{\bf x}|^{-3}$ and $T_{ij}$ 
as $|{\bf x}|^{-6}$. The decay is exponentially fast for non-zero $m$.

\section{Gauge Fields}\news
\label{sec:gauge}

In this section we derive scaling identities in a selection of field
theories with gauge fields included. Our examples have both abelian
and non-abelian gauge fields.

We start by recalling that in pure abelian gauge theory in $n$ space
dimensions the energy is
\be
E = \frac{1}{4} \int f_{ij}f_{ij} \, d^nx \,,
\ee
where $f_{ij} = \pr_i a_j - \pr_j a_i$. The ``improved'', gauge
invariant stress tensor is
\be
T_{ij} = f_{ik}f_{jk} - \frac{1}{4}\delta_{ij} f_{lm}f_{lm} \,.
\ee
The field equations, $\pr_i f_{ij} = 0$, imply that $\pr_i T_{ij}
= 0$. However, this theory has no solitons.

Scalar electrodynamics has a complex scalar field $\phi$ coupled 
to the gauge potential $a_i$. In two space dimensions, the gauge invariant 
energy is
\be
\label{electroenergy}
E = \int \left( \frac{1}{4} f_{ij}f_{ij}  
+ \half \overline{D_i\phi}D_i\phi + V \right) \,
d^2x \,.
\ee
The only non-vanishing component of $f_{ij}$ is the magnetic field
$B = f_{12}$, and $D_i\phi = \pr_i\phi - ia_i\phi$ is the covariant 
derivative of $\phi$. $V$ is a function of
$\overline{\phi} \phi$. We assume that the minimal value of $V$ is
zero. If $V$ attains this minimal value at non-zero $\phi$ then there is
spontaneous symmetry breaking, and finite energy vortex solutions 
are possible \cite{ManSut}. A vortex has the property that 
$\overline{\phi} \phi$ minimizes $V$ around the circle at infinity, 
and the phase of $\phi$ has non-trivial winding there. The winding
number $N$, being an integer, gives the vortex topological stability. 
$D_i\phi$ and $B$ both vanish exponentially fast towards infinity; however
$a_i$ has a non-zero integral around the circle at infinity, related
to the winding of $\phi$, and hence the vortex carries a magnetic
flux. The flux is $2 \pi N$ in our units.

For the energy function (\ref{electroenergy}), the stress tensor is
\be
T_{ij} = f_{ik}f_{jk} + \half\overline{D_i\phi}D_j\phi
+ \half D_i\phi \overline{D_j\phi} 
- \delta_{ij}\left( \frac{1}{4} f_{lm}f_{lm}  
+ \half\overline{D_l\phi}D_l\phi + V \right) \,.
\ee
Explicitly, the components are
\bea
T_{11} &=& \half B^2 + \half\overline{D_1\phi}D_1\phi
- \half\overline{D_2\phi}D_2\phi - V \,, \\
T_{22} &=& \half B^2 - \half\overline{D_1\phi}D_1\phi
+ \half\overline{D_2\phi}D_2\phi - V \,, \\
T_{12} &=& \half\overline{D_1\phi}D_2\phi + \half\overline{D_2\phi}D_1\phi \,.
\eea
For any vortex solution, the integral of each component of $T_{ij}$
over $\R^2$ vanishes. Working with $T_{11} + T_{22}$, $T_{11} - T_{22}$
and $T_{12}$, one obtains the identities
\bea
\label{vortid1}
&& \int \left( B^2 - 2V \right) \, d^2x = 0 \,, \\
\label{vortid2}
&& \int (\overline{D_1\phi}D_1\phi - \overline{D_2\phi}D_2\phi)
 \, d^2x = 0 \,, \\
\label{vortid3}
&& \half \int (\overline{D_1\phi}D_2\phi + \overline{D_2\phi}D_1\phi)
 \, d^2x = 0 \,.
\eea
The first of these is Derrick's theorem, and the remaining two are
perhaps less familiar identities associated with squeezing and
stretching a vortex in orthogonal directions.

In three dimensions, a non-abelian gauge theory with scalar Higgs fields can
have monopoles as soliton solutions. We consider here the standard, 
simplest example of such a theory. The gauge group is $SU(2)$ and the 
gauge potential $A_i$ is coupled to an adjoint scalar field $\Phi$, 
with no Higgs potential term. The energy is
\be
E = \int \left\{ -\frac{1}{8} \hbox{Tr}\, (F_{ij}F_{ij}) 
- \frac{1}{4} \hbox{Tr}\, (D_i\Phi D_i\Phi) \right\} \, d^3x
\ee
where $F_{ij} = \pr_iA_j - \pr_jA_i + [A_i,A_j]$ and 
$D_i\Phi = \pr_i\Phi + [A_i,\Phi]$. For details about the boundary conditions,
field equations, and the interpretation of the soliton solutions as
magnetic monopoles, see \cite{ManSut,Shn}.

The gauge invariant stress tensor is
\be
T_{ij} = -\half \hbox{Tr}\, \left( F_{ik}F_{jk} + D_i\Phi D_j\Phi \right)
+\delta_{ij}\left\{ \frac{1}{8} \hbox{Tr}\, (F_{lm}F_{lm}) 
+ \frac{1}{4} \hbox{Tr}\, (D_l\Phi D_l\Phi) \right\} \,,
\ee
which satisfies $\pr_i T_{ij} = 0$ as usual. The scaling identities
obtained from the vanishing of the integrals of each component of the
stress tensor are simplest if one works with the combinations
$T_{11} + T_{22}$, $T_{12}$, and related quantities obtained by
permuting indices. One finds
\bea
&& -\half \int \hbox{Tr}\, \left( F_{12}F_{12} 
- D_3\Phi D_3\Phi \right) \, d^3x = 0 \,, \\
&& -\half \int \hbox{Tr}\, \left( F_{13}F_{23} 
+ D_1\Phi D_2\Phi \right) \, d^3x = 0 \,,
\eea
plus permutations.

These identities are valid for any solutions of the field equations 
approaching the vacuum at infinity. Such solutions are not just the 
stable monopole and multi-monopole solutions, but also the unstable
monopole-antimonopole solutions \cite{Tau,Rub,KK}, and the various, recently
found solutions that consist of chains of monopoles, antimonopoles, 
and Higgs vortex rings \cite{KKS}. Note that for all these solutions, $F_{ij}$
and $D_i\Phi$ decay as $|{\bf x}|^{-2}$ or better for large $|{\bf x}|$, so
$T_{ij}$ decays as $|{\bf x}|^{-4}$, fast enough to avoid surface
corrections to the identities above.

The identities we have found are actually trivial for the stable
monopole solutions of minimal energy, because these satisfy the
Bogomolny equations \cite{Bo}
\be
F_{ij} = \epsilon_{ijk} D_k\Phi
\ee
(or $F_{ij} = -\epsilon_{ijk} D_k\Phi$ for antimonopoles). As has been 
noted before \cite{Loh}, the stress tensor vanishes identically if the 
Bogomolny equations are satisfied. Non-trivial identities are obtained for 
the unstable solutions mentioned above. Non-trivial identities are 
also obtained for monopoles in modified field theories where, for 
example, there is a non-vanishing Higgs potential $V(\hbox{Tr}\, \Phi^2)$ 
as part of the energy density, and hence no Bogomolny equations.

Let us return for a moment to scalar electrodynamics in two dimensions.
Here the vortices satisfy the Bogomolny equations
\bea
&& D_1\phi + iD_2\phi = 0 \,, \\
&& B - \half (1 - \overline{\phi}\phi) = 0 \,,
\eea
if the potential $V$ has the special form 
\be
V(\overline{\phi}\phi) = \frac{1}{8} (1 - \overline{\phi}\phi)^2 \,.
\ee
In this case also, the stress tensor vanishes at each point \cite{Loh}, and
the scaling identities (\ref{vortid1}) -- (\ref{vortid3}), obtained
earlier, are all trivial.

Our final example is pure Yang--Mills theory in four space dimensions,
with energy (euclidean action)
\be
E = - \frac{1}{8} \int \hbox{Tr}\, 
\left( F_{\mu\nu}F_{\mu\nu} \right) \, d^4x \,,
\ee
and stress tensor
\be
T_{\mu\nu} = -\half \hbox{Tr}\, \left( F_{\mu\sigma}F_{\nu\sigma} \right)
+\frac{1}{8} \delta_{\mu\nu} \hbox{Tr}\, \left(
F_{\sigma\tau}F_{\sigma\tau} \right) \,.
\ee
The stress tensor is traceless here. Scaling identities satisfied
by finite energy solutions of the Yang--Mills field equations 
$D_{\mu} F_{\mu\nu} = 0$ are obtained by integrating 
$T_{11} + T_{22}$, $T_{12}$, etc. They are
\bea
&& -\half \int \hbox{Tr}\, \left( F_{12}F_{12} - F_{34}F_{34} \right) 
= 0 \,, \\
&& -\half \int \hbox{Tr}\, \left( F_{13}F_{23} + F_{14}F_{24} \right) 
= 0 \,.
\eea
and the similar identities obtained by permuting indices. 

These identities rely, as usual, on there being no contribution from
surface terms. Now, since the Yang--Mills equations are conformally 
invariant, many solutions on $\R^4$, including all instantons, arise
from smooth solutions on $S^4$ through stereographic projection
(conformal decompactification). For these, the field tensor in $\R^4$
decays as $|x|^{-4}$, and the energy density and stress tensor decay
as $|x|^{-8}$, so surface terms vanish.

Instantons, which satisfy the (anti-)self-dual equations \cite{BPST}
\be
F_{\mu\nu} = \pm\half\epsilon_{\mu\nu\sigma\tau}F_{\sigma\tau} \,,
\ee
in fact satisfy the identities above trivially, because the stress tensor
vanishes everywhere. However, there are non-trivial solutions of the 
Yang--Mills equations which are not instantons \cite{SSU,SS}, and for
these, the identities are non-trivial.

\section{Conclusion}\news
\label{sec:concl}

We have established a set of scaling identities satisfied by static,
localized solutions of field theories in various dimensions. They are
proved either by considering variations of the energy function, or by
considering the conserved stress tensor. A
particular combination of these identities is Derrick's theorem. We
have presented our results using a number of examples of field
theories known to have soliton solutions, and it would be
straightforward to generalize them to other field theories.

One possible application of our results is to the improvement of 
approximate soliton solutions by suitable squeezing and stretching
transformations. We have not found an algorithm that
takes a generic field configuration, and in one step transforms it to a
field configuration that satisfies all the scaling
identities. However, we have suggested an iterative approach that
reduces the energy at each step, and could be useful if applied to
some known, approximate Skyrmion solutions.

The identities become trivial for solutions satisfying first order 
Bogomolny equations or the (anti-)self-dual Yang--Mills equations.

\subsection*{Acknowledgements}

I am grateful to Mihalis Dafermos and Gary Gibbons for discussions
about the stress tensor, and for drawing my attention to some of the
relevant literature.

\end{document}